\documentclass[aps,preprint,onecolumn,floatfix,superscriptaddress,nofootinbib]{revtex4-1}

\usepackage{ulem}
\usepackage[utf8]{inputenc}
\usepackage{amsmath}
\usepackage{amssymb}
\usepackage{graphicx,bm}
\usepackage[colorlinks,citecolor=blue]{hyperref}
\usepackage[caption=false]{subfig}
\usepackage{slashed}

\usepackage{xcolor}

\newcommand{\be}{\begin{eqnarray}}
\newcommand{\ee}{\end{eqnarray}}

\newcommand{\JCAP}{{JCAP}}

\begin{document}

\normalsize


\title{Constraints on dark matter annihilation in the Large
Magellanic Cloud from multiple low-frequency radio observations}

\author{Zhanfang Chen}
\affiliation{Institute of Modern Physics, Chinese Academy of Sciences, Lanzhou 730000, China}
\affiliation{Department of Astronomy, Xiamen University, Xiamen, Fujian 361005, China}

\author{Feng Huang}
\email{Corresponding author: fenghuang@xmu.edu.cn}
\affiliation{Department of Astronomy, Xiamen University, Xiamen, Fujian 361005, China}

\author{Taotao Fang}
\affiliation{Department of Astronomy, Xiamen University, Xiamen, Fujian 361005, China}



\begin{abstract}
Low-frequency radio emission from the Large Magellanic Cloud~(LMC) is assumed to be dominated by nonthermal synchrotron radiation from energy loss of  energetic $e^+/e^-$ in magnetic field. Two different kinds of sources of $e^+/e^-$, dark matter~(DM) annihilation and cosmic rays~(CR) related to massive stars, are taken into account in this paper. We fit the multiple low-frequency radio observations, from 19.7 MHz to 1.4 GHz, with a double power-law model $S_{\mathrm{nth}} =S_{\mathrm{DM}}(\frac {\nu}{\nu_{\star}})^{-\alpha_{DM}}+S_{\mathrm{CR}}( \frac {\nu}{\nu_{\star}})^{-\alpha_{\mathrm{CR}}} $. $\nu_{\star}$ is set to be $1.4$ GHz. For gaugino annihilation, the flux from dark matter annihilation exhibits a $\nu^{-0.75}$ power-law dependence on frequency. We fix $\alpha_{\mathrm{DM}}$ and treat $S_{\mathrm{CR}}$, $S_{\mathrm{DM}}$, and $\alpha_{\mathrm{CR}}$ as free parameters when fitting the low frequency radio data. Depending on whether thermal emission is included or not, our best-fit values for $S_{\mathrm{DM}}(\nu_{\star})$ range from  98.4 Jy to 114.8 Jy. Furthermore, we derive upper limits on the particle properties of dark matter associated with gaugino annihilation across various radio frequencies. Future low-frequency and high-resolution radio surveys are expected to serve as promising and powerful tools for constraining the properties of dark matter.
\end{abstract}

\date{\today}

\maketitle
	

\section{Introduction}
The particle nature of dark matter remains unclear, indirect detection of dark matter aims to search electromagnetic signals induced by dark matter annihilation in astronomical systems, which could put strong constraints on dark matter mass vs. annihilation cross section parameter space~( $m_{\chi}$ vs. ${\langle \sigma v \rangle}$ ) \citep{Feng:2010gw,reviewII}. In the scenario of the widely discussed dark matter candidate, Weakly Interacting Massive Particles~(WIMPs), the final products of self-annihilation include $e^+/e^-$ \citep{WIMPI,WIMPII}. Low frequency observations of nearby galaxies have potential to detect the synchrotron radiation from energy loss of such kind of energetic $e^+/e^-$ in magnetic field. Dwarf spheroidal galaxies~(dSphs), such as the satellites of our Milk Way, are among the most promising targets to detect dark matter annihilation signal for their deficiency in star formation and then lack of cosmic ray $e^+/e^-$. Several groups have conducted deep radio observations of the Local Group dSphs with the Australia Telescope Compact Array~(ATCA) in the frequency band 1.1 GHz$-$3.1 GHz \citep{ColafrancescoATCTI, ColafrancescoATCTII, ColafrancescoATCTIII, ColafrancescoATCTReti}, with the Green Bank Telescope~(GBT) at the frequency 1.4 GHz \citep{GBTI, GBTII}. Very recently, the first observational limits on diffuse synchrotron emission from 14 dSphs at radio frequency below 1.0 GHz have been presented by the combination of
survey data from the Murchison Wide field Array (MWA) and the Giant Metre-wave Radio Telescope(GMRT) \citep{MWA2019}. Additionally, similar studies have been conducted with data from the TIFR-GMRT sky survey at the frequency 150 MHz for a stacking analysis of 23 dwarf spheroidal galaxies and with FAST data for the Coma Berenices dwarf galaxy\citep{Basu2021,Guo2022}. However, no significant detection of a diffuse radio continuum found in these observations. Such a kind of non-detection could also put upper limits on dark matter annihilation, while the limits vary widely due to the large uncertainty of assumed magnetic field and other related astrophysical properties. The nearest dwarf irregular galaxies (dIrrs),  the Large Magellanic Cloud (LMC), is also an optimal target for synchrotron radiation searches from dark matter annihilation because of its proximity and its large mass-to-light ratio \citep{Sofue1999,Gammaldi:2021zdm}. Meanwhile, photometric and spectral measurements of LMC have covered almost the entire electromagnetic spectrum, which enables us to deduce its properties of magnetic field and star formation. 

Recently, Regis et al. used the Australian Square Kilometer Array Pathfinder (ASKAP) to observe the LMC at 888 MHz but found no clear dark matter signal, only setting an upper limit on annihilation\citep{Regis2021}. A subsequent study has been carried out by Siffert et al., however, they considered constraints from radio emission from 1.4 GHz and 4.8GHz, which may contain large fraction of thermal free$\--$free emission from ionized hydrogen clouds \citep{Siffert2011}. Pioneering radio observations of LMC date back to 1950s \citep{Shain1959, Mills1959, Alvarez1987, Klein1989}, which include the lowest frequency 19.7 MHz. For et al. present the first low-frequency (the lowest frequency is 76 MHz) MWA radio continuum flux of LMC from the GaLactic Extragalactic All-Sky MWA (GLEAM) survey \citep{For2018}. Both Tasitsiomi et al. \citep{Tasitsiomi2004} and Chan et al. \citep{Chan2022} utilized the radio continuum observations of the LMC by Haynel et al. in 1991 \citep{Haynes1991}, covering the frequency range from 19.7 MHz to 8.55 GHz, to constrain the properties of dark matter. Tasitsiomi et al. considered the observed flux as an upper limit resulting from dark matter annihilation, whereas Chan et al. required an extremely suppressed cosmic-ray flux to achieve good agreement with the observations. However, unlike in dwarf spheroidal galaxies, the low-frequency radio emission from the LMC is generally believed to be dominated by cosmic rays. This paper aims to revisit the constraints on dark matter annihilation in the LMC, focusing on updated low-frequency radio observations. 
We fit the multiple low-frequency radio observations, from 19.7 MHz to 1.4 GHz, with a double power-law model $S_{\mathrm{nth}} =S_{\mathrm{DM}}(\frac {\nu}{\nu_{\star}})^{-\alpha_{DM}}+S_{\mathrm{CR}}( \frac {\nu}{\nu_{\star}})^{-\alpha_{\mathrm{CR}}}$. $\nu_{\star}$ is set to be $1.4$ GHz. In the study of \citet{Tasitsiomi2004}, 
they demonstrated that the synchrotron radiation from dark matter annihilation scales as $\nu^{-0.75}$ when a specific annihilation channel, such as gaugino annihilation, is considered. Following this approach, we fix $\alpha_{\mathrm{DM}}$ while treating $S_{\mathrm{CR}}$, $S_{\mathrm{DM}}$, and $\alpha_{\mathrm{CR}}$ as free parameters. Using the Markov Chain Monte Carlo (MCMC) method, we then perform the data fitting to derive upper limits on the synchrotron emission induced by dark matter annihilation at different radio frequencies.

In this paper, we present revised constraints on dark matter in LMC with updated multiple low-frequency radio observations included.  In Section II, we fit the multiple low-frequency radio emissions with a double power-law model and present the upper limits of synchrotron emission induced by dark matter annihilation at different radio frequencies. In Section III, we describe our astrophysical model for LMC, including dark matter distribution, magnetic field, and properties related to energetic $e^+/e^-$ transportation. We then derive the synchrotron emission induced by dark matter annihilation and constraints on the particle parameter space of dark matter. Lastly, we present our conclusions and discussion in Section IV.

\section{Radio data and model fitting}
\subsection{Radio data}
Radio data from 19.7 MHz to 1.4 GHz are taken from Table 2. in \citet{For2018} as illustrated here in Table \ref{tab:integflux}. The study presented the overall radio continuum morphology between 76 MHz and 227 MHz from the GLEAM survey conducted by MWA\cite{For2018}. The integrated flux densities $S_{\nu}$ in Table 2. in \citet{For2018} are extracted from $8^{\circ}\times8^{\circ}$ images centred on the LMC. This work alone has added 20 data points below 1 GHz which enable us to do a statistical data fitting.
\begin{table}
\begin{center}
\begin{tabular}{rcccc}
 \hline
   $\nu$ (MHz) & flux (Jy) & reference \\ \hline
  19.7 & $5270\pm1054$ & \cite{Shain1959} \\
  45   & $2997\pm450$ & \cite{Alvarez1987} \\
  85.5 & $3689\pm400$ & \cite{Mills1959} \\
  98.6 & $2839\pm600$ & \cite{Mills1959} \\
  158  & $1736\pm490$  & \cite{Mills1959} \\
  76   & $1855.3\pm315.6$  & \cite{For2018} \\
  84   & $1775.7\pm302 $  &  \cite{For2018} \\
  92   & $1574.9\pm267.9$ &  \cite{For2018} \\
  99   & $1451.6\pm247  $ &  \cite{For2018} \\
  107  & $1827.6\pm310.8$  &  \cite{For2018} \\
  115  & $1627.5\pm276.8$  &  \cite{For2018} \\
 123  & $1643.3\pm279.4$  &  \cite{For2018} \\
 130   & $1571.8\pm267.3$  &  \cite{For2018} \\
 143  & $1663.7\pm282.9$  &  \cite{For2018} \\
150   & $1450.1\pm246.6$  &  \cite{For2018} \\
158  & $1350.4\pm229.6$  &  \cite{For2018} \\
166  & $1204.3\pm204.8$  &  \cite{For2018} \\
174  & $1341.4\pm228.1$  &  \cite{For2018} \\
181  & $1247.1\pm212.1$  &  \cite{For2018} \\
189  & $1223.9\pm208.1$  &  \cite{For2018} \\
197  & $1109.8\pm188.8$  &  \cite{For2018} \\
204  & $1235.3\pm210.1$  &  \cite{For2018} \\
212  & $1121.6\pm190.8$  &  \cite{For2018} \\
219  & $1032.4\pm175.6$  &  \cite{For2018} \\
277  & $1019.7\pm173.4$  &  \cite{For2018} \\
408  & $925\pm30$  &   \cite{Klein1989} \\
1400  & $384\pm30$ &    \cite{For2018} \\
1400   & $529\pm30$ &    \cite{Klein1989} \\
 \hline
\end{tabular}
\caption{ Updated multiple low-frequency radio data }
\label{tab:integflux}
\end{center}
\end{table}

\subsection{Model fitting}

%

We consider two different sources of $e^+/e^-$ that contribute to non-thermal radiation: dark matter (DM) annihilation and cosmic rays associated with massive stars. The internal mechanisms responsible for producing $e^+/e^-$ differ significantly, leading to distinct $e^+/e^-$ spectra and, consequently, different synchrotron radiation power-law indices. Therefore, we adopt a double power-law model to fit the data.
\begin{equation}
\label{eq1}
S_{\mathrm{nth}} =S_{\mathrm{DM}}( \frac {\nu}{\nu_{\star}})^{-\alpha_{\mathrm{DM}}}+S_{\mathrm{CR}}( \frac {\nu}{\nu_{\star}})^{-\alpha_{\mathrm{CR}}},
\end{equation}
where $\nu$ is the radio frequency.
 $S_{\mathrm{DM}}$ ($S_{\mathrm{CR}}$) represents radio flux from dark matter annihilation (cosmic rays) contribution at $\nu_{\star}$ and $S_{\mathrm{nth}}$ is the non-thermal radiation. 

 

The spectrum of dark matter annihilation flux exhibits a complex dependence on frequency that varies across different annihilation channels. In our study, we consider gaugino annihilation, which mainly produces fermion pairs directly through $\chi\chi \rightarrow ff$, with cross sections scaling as $M_f^2$ for different fermions. Below $1.4\,\mathrm{GHz}$, the flux $F \propto \frac{dN_e}{dE}$ follows an approximate power-law dependence on frequency, as shown in Fig.5 of Tasitsiomi et al.\cite{Tasitsiomi2004}. The relationship between $\frac{dN_e}{dE}$ and frequency can be described by:
\begin{equation}
\frac{dN_e}{dE} \propto 2.48 - 1.22 x^{-0.5} + 0.22 x^{2} + 0.10 x^{-1.5} - 1.54 x^{0.5} - 0.04 x^{3}
\end{equation}
where $x = \frac{E_e}{m_\chi c^2} \simeq \frac{4.94 \times 10^{-4} B_\mu^{-0.5} \nu^{0.5}}{m_\chi}$.The electron energies responsible for the maximum synchrotron emission below $1.4$ GHz are relatively low. Tasitsiomi et al. demonstrated that at low electron energies, both the flux and frequency scaling follow a $\nu^{-0.75}$ power-law dependence \cite{Tasitsiomi2004}.
Then, the power-law model can be rewritten as: 
\begin{equation}
\label{eq3}
S_{\mathrm{nth}} =S_{\mathrm{DM}}( \frac {\nu}{\nu_{\star}})^{-0.75} +S_{\mathrm{CR}}( \frac {\nu}{\nu_{\star}})^{-\alpha_{\mathrm{CR}}}.
\end{equation}

The parameters $S_{\mathrm{CR}}$ and $\alpha_{\mathrm{CR}}$ are generally associated with synchrotron emission from cosmic ray $e^+e^-$ pairs. A typical value of $\alpha_{\mathrm{CR}}$ adopted in the literature is around 0.80 \citep{Condon1992}. However, \citet{For2018} found a best-fit value of $\alpha_{\mathrm{CR}} = 0.55$ when fitting radio data in the frequency range from 19.7 MHz to 8.55 GHz in LMC. Therefore, we treat $\alpha_{\mathrm{CR}}$ as a free parameter to be constrained by data fitting. The normalization factor $S_{\mathrm{CR}}$ also remains uncertain, as it can vary depending on the model and data set used. From a theoretical perspective, supernova remnants (SNRs) are considered the primary sources of cosmic rays, and their contribution can be inferred from the star formation rate \citep{Berezhko2014}. Infrared (IR) emission is predominantly thermal radiation from dust heated by ultraviolet (UV) light emitted by young, massive stars. These same stars eventually evolve into supernovae, whose remnants are believed to produce high-energy $e^+e^-$ pairs. A well-established correlation exists between synchrotron and IR emission in normal galaxies \citep{Rieke2009}. The empirical relationship between radio and infrared luminosity is as follows:
\begin{equation}
\label{eq2}
{\rm log} ~\frac {L_{\mathrm{CR}}}{ W \cdot \mathrm{Hz}^{-1}} = 1.032 \times {\rm log}~\frac {L_{24\mu m}}{ L_{\bigodot}} + 11.642
\end{equation}

If this empirical correlation holds strictly in the LMC, the 1.4 GHz radio emission due to cosmic rays can be estimated from the precisely measured infrared luminosity. For instance, we adopt $S_{24\mu \mathrm{m}} = 8080 \pm 320$ Jy, as reported by \citet{Lawton2010}. At a distance of 50.1 kpc, this corresponds to a cosmic-ray-induced radio flux of $S_{\mathrm{CR}}(1.4,\mathrm{GHz}) = 177.9 \pm 7.2$ Jy. However, since this empirical correlation is derived from normal galaxies, it carries substantial uncertainties, and the validity of the inferred $S_{\mathrm{CR}}(1.4,\mathrm{GHz})$ for the LMC remains uncertain. Therefore, in this study, we also treat $S_{\mathrm{CR}}(1.4,\mathrm{GHz})$ as a free parameter. 


As discussed above, we fix $\alpha_{\mathrm{DM}}$ while treating $S_{\mathrm{CR}}$, $S_{\mathrm{DM}}$, and $\alpha_{\mathrm{CR}}$ as free parameters. Using the Markov Chain Monte Carlo (MCMC) method, we then perform the data fitting to derive upper limits on the synchrotron emission induced by dark matter annihilation at different radio frequencies. We make use of the {\it emcee} code to do MCMC sampling, which is available online \footnote {\it https://emcee.readthedocs.io/en/stable/}. The script '{\it Corner.py}' is used for plotting the contours of MCMC samplings \citep{Foreman2016}. The logarithmic likelihood function is defined as:
\begin{equation}\label{S}
  {\rm ln}~p(S_{\mathrm{nth}}|\alpha_{\mathrm{CR}},S_{\mathrm{CR}},S_{\mathrm{DM}}) = -\frac{1}{2}\sum\limits_{i=1}^N\frac{(S_i-S_{{\rm model},i})^2}{\sigma_i^2},
\end{equation}

where $N$ is the total number of observational data, $S_i$ is the observed flux in different frequencies. $S_{{\rm model},i}$  is the theoretical calculation of flux, which is expressed by $S_{\mathrm{nth}}$ in Eq.\ref{eq3}.
Since calculating the theoretical flux is very time-consuming, we calculate the flux under different parameters and obtain the flux of any parameter through interpolation. If the $\alpha_{\mathrm{CR}}$ and $S_{\mathrm{CR}}$ values are set as fixed parameters, we can identify the contributions corresponding to dark matter annihilation contributions at different frequencies. We further obtain the flux at any set of the three fitting parameters by interpolation.

\begin{figure*}
\begin{minipage}[h]{0.39\linewidth}
\center{\includegraphics[width=1\linewidth]{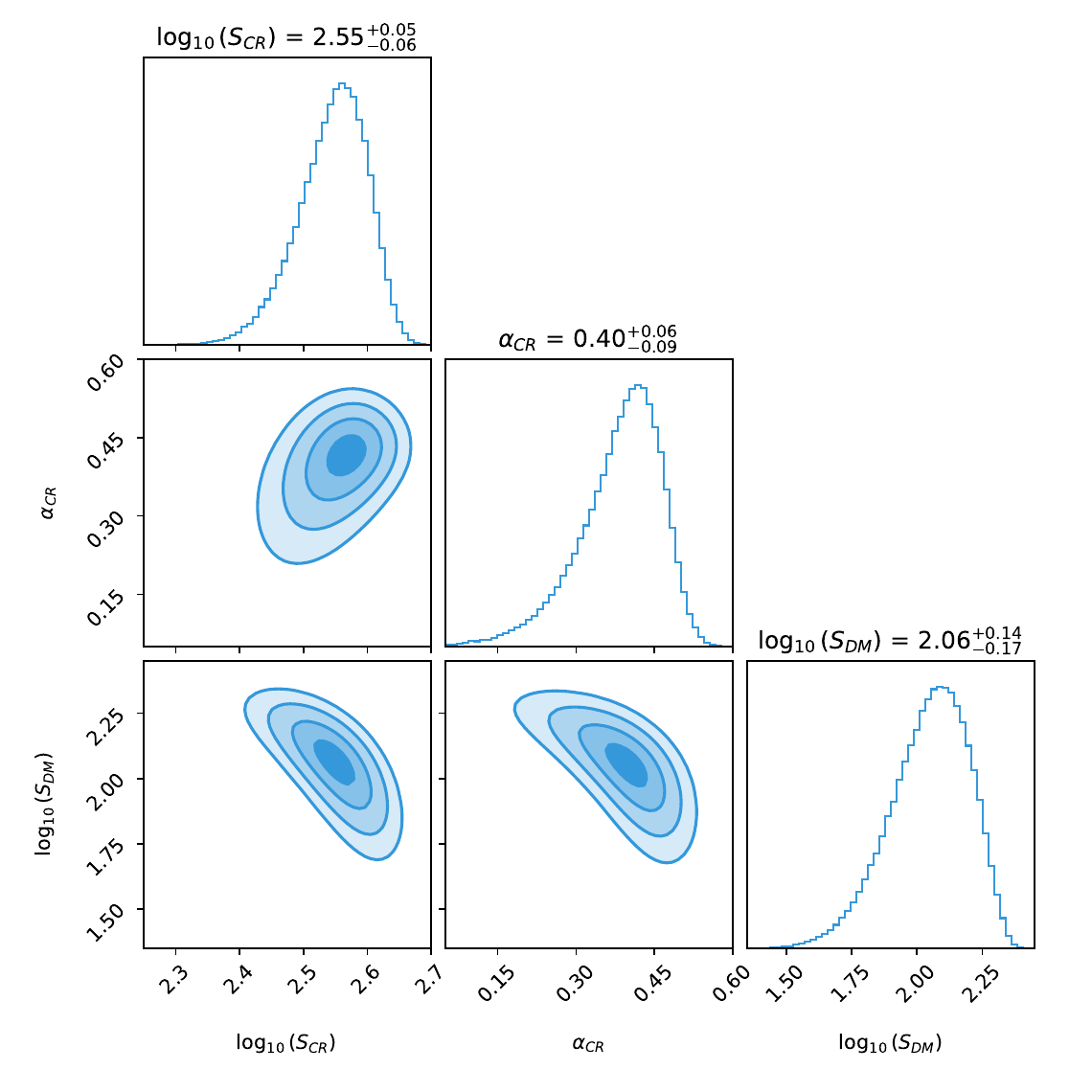} \\ }
\end{minipage}
\hfill
\begin{minipage}[h]{0.59\linewidth}
\center{\includegraphics[width=1\linewidth]{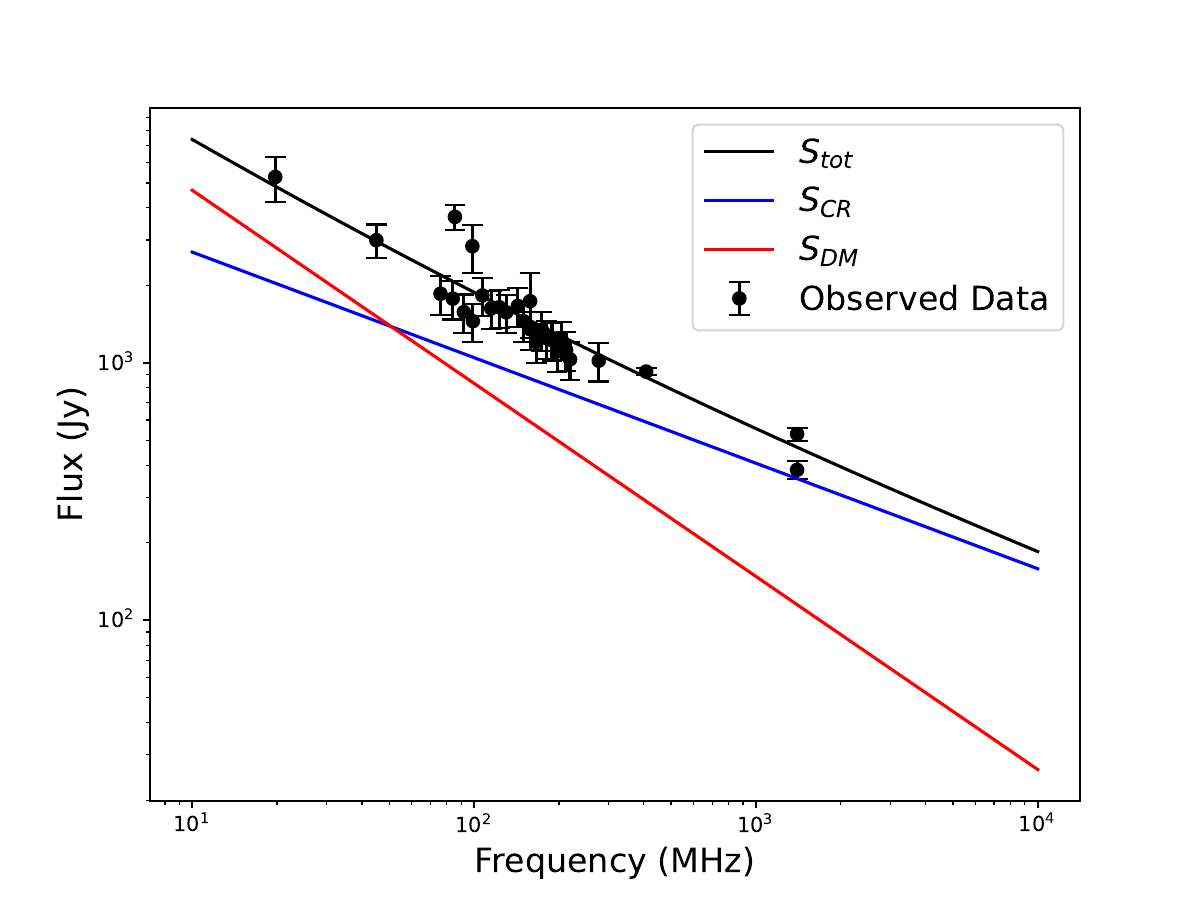} \\ }
\end{minipage}
\caption{The left figure shows the contour of the probability distribution of radio flux with three free parameters. During the fitting process, the thermal radiation flux was set to zero. The right figure shows the radio flux vs. frequency $\nu$. The scatter points represent the observed data listed in Table \ref{tab:integflux}. The blue line represents synchrotron emission from CRs, the red line represents the upper limit on synchrotron emission associated with DM annihilation . The black line is the sum of these two components. }
\label{non-thermal}
\end{figure*}

We present the contours of the probability distribution of flux with three free parameters ($\alpha_{\mathrm{CR}}$, ${\rm log_{10}}S_{\mathrm{CR}}$, and ${\rm log_{10}}S_{\mathrm{DM}}$) for a given value of $\alpha_{\mathrm{DM}}$, shown in the left panel of Fig.~\ref{non-thermal}. When fitting the observational data using only non-thermal radiation, the best-fit parameters are $\alpha_{\mathrm{CR}} = 0.40_{-0.06}^{+0.09}$, ${\rm log_{10}} S_{\mathrm{CR}} = 2.55_{-0.06}^{+0.05}$, and ${\rm log_{10}}S_{\mathrm{DM}} = 2.06_{-0.14}^{+0.17}$. 
From these results, the synchrotron emission from cosmic rays follows:
$S_{\mathrm{CR}} = 354.8(\nu/\rm 1.4\,GHz)^{-0.4}$ and the upper limit for dark matter annihilation-related synchrotron emission is: $S_{\mathrm{DM}} = 114.8 (\nu/\rm 1.4\,GHz)^{-0.75}$. 
We then present the two-component fit to the multi-frequency low-frequency radio data of the LMC in the right panel of Fig.~\ref{non-thermal}. The result shows that the upper limit of the dark matter flux exceeds the cosmic ray flux at frequencies below 50 MHz, but falls below it at higher frequencies.

\begin{figure*}
\begin{minipage}[h]{0.39\linewidth}
\center{\includegraphics[width=1\linewidth]{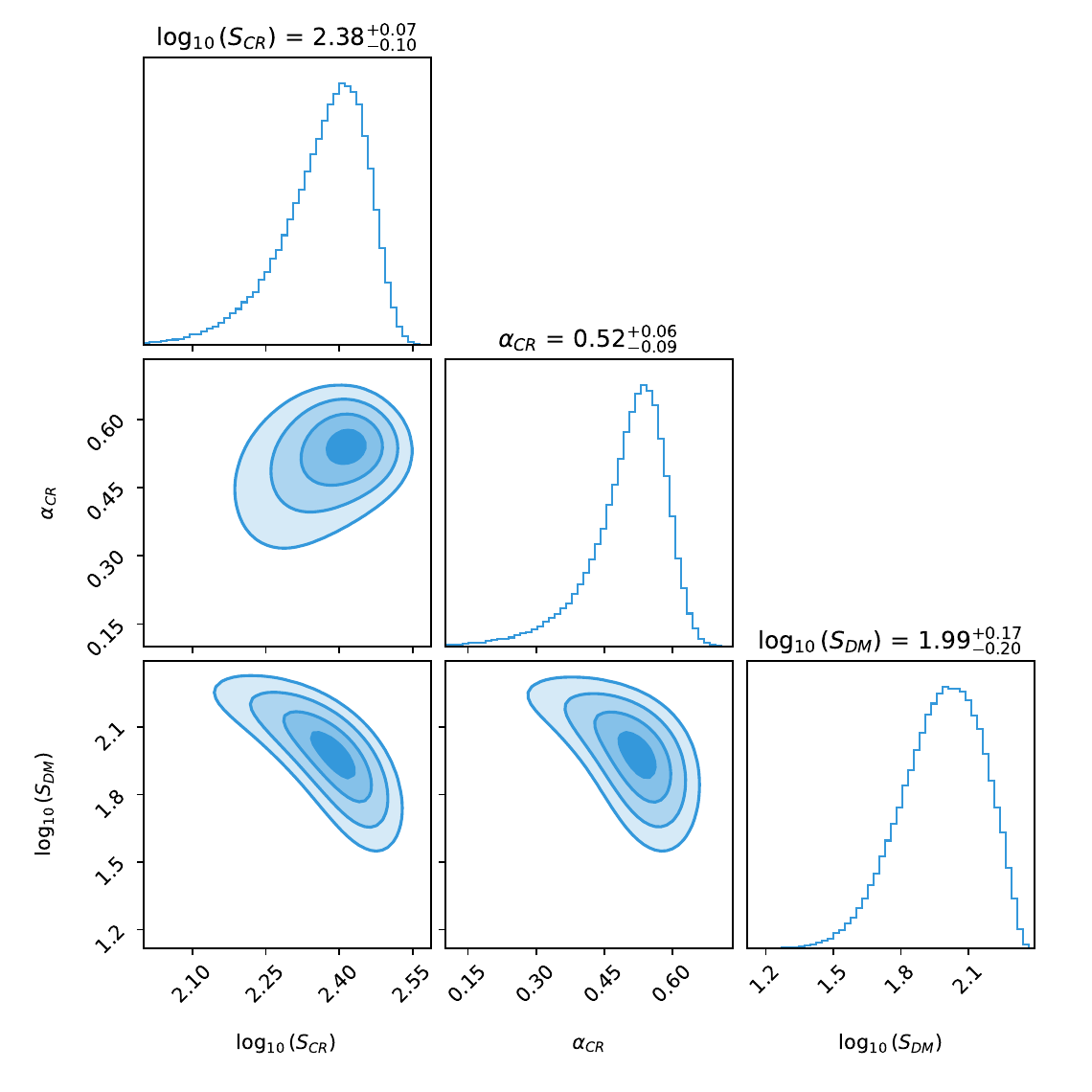} \\ }
\end{minipage}
\hfill
\begin{minipage}[h]{0.59\linewidth}
\center{\includegraphics[width=1\linewidth]{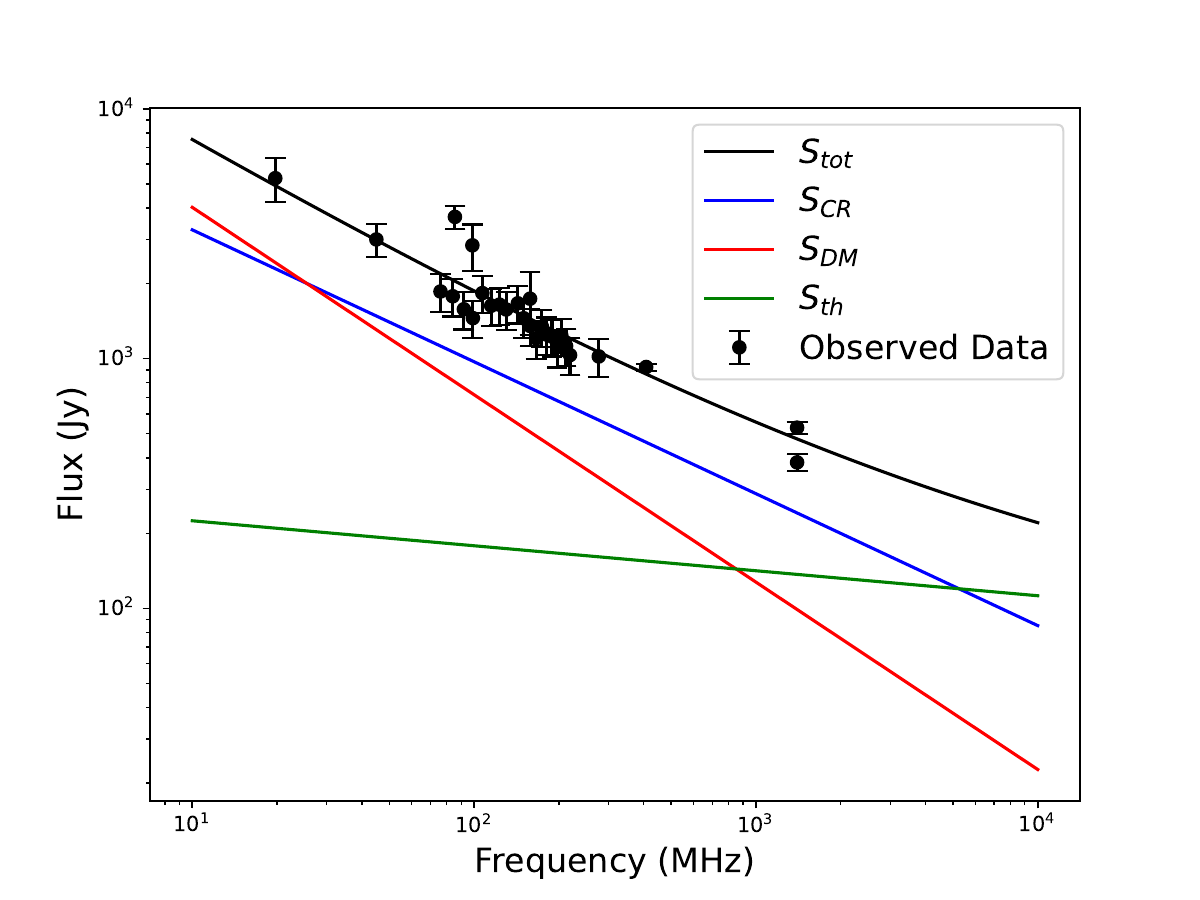} \\ }
\end{minipage}
\caption{The left figure shows the contour of the probability distribution of radio flux with three free parameters. The radio flux consists of three components: dark matter flux, cosmic ray flux, and thermal radiation flux.  The right figure shows the radio flux vs. frequency $\nu$. The scatter points represent the observed data listed in Table \ref{tab:integflux}. The blue line is synchrotron emission from CRs, the red line represents the allowed upper limit on synchrotron emission related to DM annihilation and the green line corresponds to thermal emission. The black line is the sum of these two components.}
\label{themal}
\end{figure*}

At frequencies below 1.4 GHz, previous studies have shown that the total radio flux $S_{\text{tot}}$ is primarily dominated by non-thermal emission $S_{\text{nth}}$ \citep{For2018}. However, the thermal component may not drop entirely to zero. \citet{Hassani2022} reported that thermal emission $S_{\text{th}}$ contributes approximately 136.8 Jy at 1.4 GHz, accounting for about 30$\%$ of the total radio flux. In the optically thin regime, thermal free–free emission from thermal electrons is typically characterized by a spectral index of $\alpha_{\mathrm{th}} = 0.1$ \citep{Klein1989}. The thermal component can therefore be modeled as:
\begin{equation} S_{\mathrm{th}} = 136.8 \left( \frac{\nu}{\nu_{\star}} \right)^{-0.1}, \end{equation}
where $\nu_{\star}$ denotes the reference frequency, taken to be 1.4 GHz. In addition to the scenario where the radio emission is assumed to originate solely from non-thermal processes (i.e., with $S_{\text{th}} = 0$), we also perform a fit that includes a thermal component following the relation $S_{\text{th}} \propto \nu^{-0.1}$. As shown in left panel in Fig.\ref{themal}, the best-fit parameters are $\alpha_{\mathrm{CR}} = 0.52_{-0.06}^{+0.09}$, ${\rm log_{10}} S_{\mathrm{CR}} = 2.38_{-0.10}^{+0.07}$, and ${\rm log_{10}} S_{\mathrm{DM}} = 1.99_{-0.17}^{+0.20}$. 
From these results, we obtain the spectral components: cosmic ray synchrotron emission $S_{\mathrm{CR}} = 239.9(\nu/\rm 1.4\,GHz)^{-0.52}$ and dark matter annihilation synchrotron emission $S_{\mathrm{DM}} = 98.4 (\nu/\rm 1.4\,GHz)^{-0.75}$.

The normalization factors for both the CR and DM components are found to decrease slightly when thermal radiation is included, which is consistent with the presence of a small but non-zero additional emission component. This adjustment reflects a more accurate decomposition of the total radio flux into thermal and non-thermal contributions. Correspondingly, the best-fit value of the synchrotron spectral index $\alpha_{\mathrm{CR}}$ increases from 0.40 (in the case without thermal radiation) to 0.52 when thermal emission is taken into account. Although both values remain lower than the canonical value of $\alpha_{\mathrm{CR}} \approx 0.80$ commonly adopted in the literature, they are in good agreement with the findings of \citet{For2018}. In their analysis of the LMC's radio emission, a double power-law model incorporating both thermal and non-thermal components was fitted to data ranging from 19.7 MHz to 8.55 GHz, yielding a best-fit value of $\alpha_{\mathrm{CR}} = 0.55$. This consistency supports the reliability of our modeling approach and suggests that the radio spectrum of the LMC may be flatter than typically assumed for normal galaxies. Moreover, the fitted values $S_{\mathrm{CR}}(1.4\,\mathrm{GHz}) = 354.8$ Jy (in the case without a thermal component) and $S_{\mathrm{CR}}(1.4\,\mathrm{GHz}) = 239.9$ Jy (in the case with a thermal component) are both higher than the value derived from the radio/far-infrared relation, $S_{\mathrm{CR}}(1.4\,\mathrm{GHz}) = 177.9\,\mathrm{Jy}$. This, in turn, provides indirect support for the argument that the empirical correlation established for normal galaxies may not apply to dwarf irregular galaxies such as the LMC directly.

The right panel of Fig.~\ref{themal} presents our three-component fit to the multi-frequency radio data from the LMC. The green line represents the thermal emission, which contributes only a small fraction to the total radio flux at frequencies below 1.4 GHz. The upper limit on the DM-induced flux (red line) exceeds the CR flux (blue line) at frequencies below 25 MHz but falls below it at higher frequencies. This is consistent with the case without the thermal component, except that the turnover frequency is around 50 MHz instead of 25 MHz. In both cases, the CR component dominates over the DM component across most of the low-frequency range considered here.


To summarize, we perform a three-parameter fit using $\alpha_{\mathrm{CR}}$, $\log_{10}S_{\mathrm{CR}}$, and $\log_{10}S_{\mathrm{DM}}$. The resulting upper limits for dark matter synchrotron emission are given by $S_{\mathrm{DM}} = 114.8(\nu/1.4,\mathrm{GHz})^{-0.75}$ (excluding thermal radiation) and $S_{\mathrm{DM}} = 98.4(\nu/1.4,\mathrm{GHz})^{-0.75}$ (including thermal radiation). In the next section, we will use this expression to derive constraints on the dark matter particle parameter space.

\section{Constraints on dark matter particle parameter space}
In the above section, we have deduced synchrotron emission from dark matter annihilation. In order to get the detailed constraints on dark matter particle properties, we firstly need information about the source function of $e^+/e^-$ produced by dark matter annihilation in LMC.

\subsection{ $e^+/e^-$ produced from DM annihilation}
The source function is the number density distribution of $e^+/e^-$ produced by DM annihilation.
\begin{equation}\label{q}
  q_e(E,r) = \frac{1}{2}\left(\frac{\rho_{\scriptscriptstyle DM}(r)}{m_{\chi}}\right)^2 \langle \sigma v \rangle \frac{dN_e}{dE},
\end{equation}
where $\langle \sigma v \rangle $ denotes the velocity averaged annihilation cross section and $m_{\chi} $ is the mass of dark matter particle. $dN_e/dE$ is the electron spectrum of dark matter annihilation. 



\begin{figure*}
\begin{minipage}[h]{0.49\linewidth}
\center{\includegraphics[width=1 \linewidth]{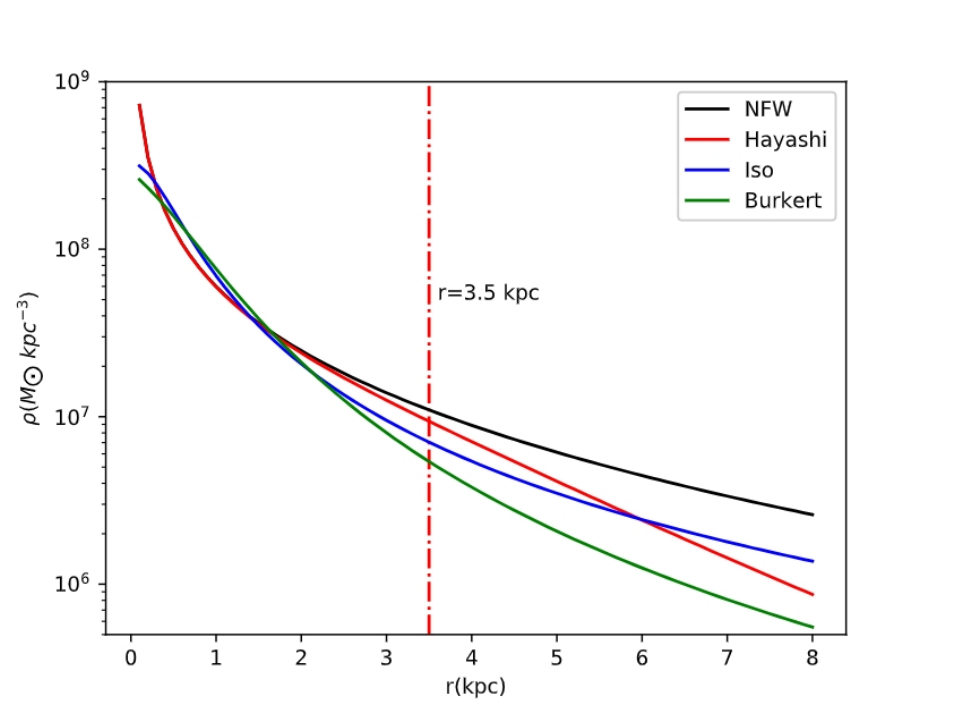} \\ }
\end{minipage}
\hfill
\begin{minipage}[h]{0.49\linewidth}
\center{\includegraphics[width=1 \linewidth]{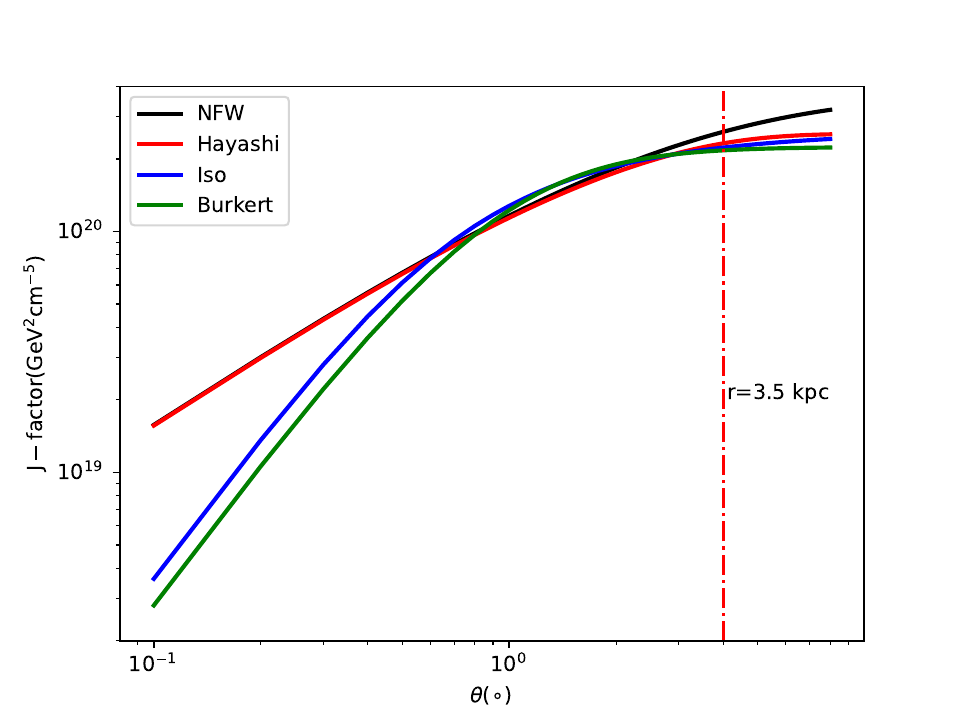} \\ }
\end{minipage}

\caption{The left figure depicts the relationship between dark matter distribution and radius, while the right figure illustrates the correlation between the J-factor and the observation angle. We consider four distinct dark matter density profiles: Hayashi (red), NFW (black), Isothermal Sphere (blue), and Burkert (green).}

\label{LMCrho}
\end{figure*}

The signal strength resulting from dark matter annihilation is directly proportional to the square of the dark matter density ($\rho_{\scriptscriptstyle DM}(r)$). When compared to dwarf spheroidal galaxies, the LMC provides more precise and comprehensive rotational data, facilitating the determination of the dark matter density profiles. However, given our limited understanding of dark matter physics on smaller scales, it is imperative to examine a range of density profiles. In Fig.\ref{LMCrho} (left), we present the density profiles, including NFW, Hayashi, Isothermal sphere, and Burkert\cite{Navarro1995iw,Moore1999gc,Salucci2000ps,Hayashi2002qv}. The associated parameters for these profiles are detailed in Table 2 of the study \citep{Siffert2011}. Additionally, the J-factor represents a crucial parameter of the dark matter density distribution for regions of interest.  The J-factor is computed by integrating the square of the dark matter density ($\rho_{\scriptscriptstyle DM}(r)$) along the line of sight while considering the astrophysical characteristics of the region  \citep{{Bonnivard2015}}:

    \begin{equation}
            J = \int \int \rho^2_\text{DM}(r) \text{d}\ell \text{d}\Omega,
        \label{eqn:jfactor}
    \end{equation}

 The right panel of Fig.\ref{LMCrho} illustrates  the values of the J-factor within the 3.5 kpc radius for four density profiles: $\rm 2.59 \times 10^{20} ~ GeV^{2}cm^{-5}$ (NFW), $\rm 2.32 \times 10^{20} ~ GeV^{2}cm^{-5}$ (Hayashi), $\rm 2.23 \times 10^{20} ~ GeV^{2}cm^{-5}$ (isothermal sphere), and $\rm 2.17 \times 10^{20}  ~GeV^{2}cm^{-5}$ (Burkert).   On the whole, at scales around 3.5 kpc, the density profiles has a negligible impact on the overall flux. Due to the similarity in J-factor values, the NFW profile is predominantly utilized in our subsequent analyses.

\subsection{$e^\pm$ propagation in LMC}

Dark matter annihilation injects electrons and positrons in the
galaxy halo at a constant rate. The propagation of these charged
particles in the tangled magnetic field can be modeled as
diffusion. They also lose energy by radiation during this process.
As a result, the $e^\pm$ spectrum satisfies the following
transport equation \citep{diffuse}:
\begin{equation}
  \frac{\partial}{\partial t}\frac{dn_e}{dE_e} =
  \nabla \left[ D(\bm{r},E) \nabla\frac{dn_e}{dE_e}\right] +
  \frac{\partial}{\partial E} \left[ b(\bm{r}, E) \frac{dn_e}{dE_e}\right]+
  q_e(E,r) \;,
\label{diffeq}
\end{equation}
where $dn_{e}/dE_{e}$ is the number density of $e^{\pm}$ per unit
energy interval, $ D(\bm{r},E)$ is the diffusion coefficient,
$b(\bm{r}, E)=dE_e/dt$ represents for the energy loss rate. For simplicity, we assume that
$D$ and $b$ are independent of spatial location. The diffusion
coefficient $D(E)$ is assumed to have a power law dependence on
energy $E$ and magnetic field $B$:
\begin{equation}
D(E) = D_0 \left(\frac{E}{E_0}\right)^{\delta} \label{D0},
\end{equation}
with $\delta$ taken as $0.3$, although it is not
clear to what extent this relation could be extrapolated. The scale of uniformity for the magnetic field in LMC should be larger than that in dSphs and smaller than that in Milky Way.  Analyses of the B/C ratio data from the Milky Way suggest that the diffusion coefficient $D_0 = 3 \times 10^{28}$~cm$^{2}$~s$^{-1}$ for the Milky Way with a range from $10^{27}$ to $10^{29}~{\rm cm}^2 s^{-1}$~\cite{Maurin:2001sj,Webber:1992dks,diffuse}. Given that the gravitational potential well of the LMC is weaker than the Milky Way's, the diffusion coefficient decreases by one to two orders of magnitude. Consequently, 
we set $D_0 = 3.0 \times
10^{27}$~cm$^{2}$~s$^{-1}$ as our benchmark values.

The energy loss rate is
\begin{equation}\label{b_loss}
	b((\bm{r},E)=b_{\rm ICS}(E)+b_{\rm syn}((\bm{r},E)+b_{\rm ion}(E),
\end{equation}
\begin{equation}\label{b_ICS}
	b_{ICS} = 0.76\frac{U_{ph}}{1 \mbox{ eV$\cdot$cm}^{-3}} \left(\frac{E}{1 \mbox{ GeV}}\right)^2,
\end{equation}
\begin{equation}\label{b_sync}
	b_{syn} = 0.025\left(\frac{B(\bm{r})}{1 \mbox{ $\mu$G}}\right)^2\left(\frac{E}{1 \mbox{ GeV}}\right)^2,
\end{equation}
\begin{equation}\label{b_ion}
	b_{ion} = 0.2\times\frac{N_{\rm H}}{1\rm cm^{-3}}[\ln(\Gamma)+6.6],
\end{equation} 
 where $N_{H} \sim 1.3 \times 10^{-6} cm^{-3}$denotes the number
density of neutral gas in LMC,  $\beta$ and $\Gamma$ represent the velocity of $e^+/e^-$ and the Lorentz factor respectively $U_{ph}$ is the total energy density of the Interstellar Radiation Field (ISRF). We use the fixed value $U_{ph} = 0.539 $ eV$\cdot$cm$^{-3}$, which is the local value of the ISRF in LMC (\cite{Weingartner2001}).

 The magnetic field model in this work follows the exponential model \cite{McDaniel2017}:
\begin{eqnarray}
B(r) = B_{0} \exp(-r/r_{c}),
\label{magnetic_field}
\end{eqnarray}
where $B_0$ represents the central magnetic field strength and .  We adopt the radius of diffusion zone as the target center radius $r_c$. In this study, we set $B_0=5 ~\mu G$ as our benchmark value.

In smaller systems, such as the Milky Way and dwarf spheroidal galaxies, we must take into account the diffusion loss of energetic $e^+/e^-$ \citep{McDaniel2017,Cholis2015}. The gravitational potential well and magnetic field strength of the LMC are relatively weaker compared to the Milky Way. When the diffusion length is comparable to the object size, the spatial morphology of electrons at equilibrium can be significantly different from the one at injection and that a significant fraction of them can escape the object, leading to a reduced radiative power. Therefore, we must consider the diffusion effect to obtain a more accurate representation of the behavior of energetic $e^+/e^-$ in small systems like the LMC. 

In Ref.~\cite{McDaniel2017,diffuse}, an analytic solution to Eq.\ref{diffeq} has been derived in the case of a spherically symmetric
system. Considering the time-independent source and the limit for
an electron number density that has already reached equilibrium,
the solution takes the form:
\begin{equation}
\frac{dn_e}{dE}\left(\bm{r},E \right) =  \frac{1}{b(\bm{r},E)} \int_E^{M_\chi}
dE' \; \widehat{G}\left(r,\Delta v \right) q_e(\bm{r},E')
\label{eq:full2}
\end{equation}
with:
\begin{eqnarray}
\widehat{G}\left(r, \Delta v\right) & = & \frac{1}{[4\pi(\Delta
v)]^{1/2}} \sum_{n=-\infty}^{+\infty} (-1)^n \int_0^{r_h} dr'
\frac{r'}{r_n}  \\ \nonumber & &
\times \left[\exp{\left(-\frac{(r'-r_n)^2}{4\,\Delta v}\right)}-
\exp{\left(-\frac{(r'+r_n)^2}{4\,\Delta v}\right)}\right]\\ \nonumber & &
\times \frac{\rho^2(r')}{\rho^2(r)}\;,\label{eq:rescaling}
\end{eqnarray}
where $r_n = (-1)^n r + 2 n r_h$ is the location of nth charge
image and $r_h$ is the radius of diffusion zone at which a free
escape boundary condition is imposed. The value of $r_h$ is
generally adopted as twice of the radius of the stellar component. Here in order to compare with observations, we also choose $r_h$ as $3.5$ kpc. The compound parameter $ \Delta v \equiv v(E) - v(E') $ encodes both diffusion and energy loss processes through:
\begin{equation}
v(E) = \int_{M_\chi}^E d\tilde{E} \, \frac{D(\tilde{E})}{b(\tilde{E})},
\label{eq:v_E}
\end{equation}
where $ \sqrt{\Delta v} $, having dimensions of length, represents the mean propagation distance of electrons during energy loss from $ E' $ to $ E $.  To enable an analytical solution of the steady-state Green’s function for the diffusion equation, spatial homogeneity of the magnetic field is required. Therefore, when calculating $ \sqrt{\Delta v} $, we apply a spatial averaging of the magnetic field.  Here, the synchrotron energy loss rate $b(\tilde{E})$ in Eq.~\ref{eq:v_E} is spatially independent and defined as
\begin{equation}
b(\tilde{E}) \approx 0.025\left(\frac{B_{\text{avg}}}{1 \mbox{ $\mu$G}}\right)^2\left(\frac{\tilde{E}}{1 \mbox{ GeV}}\right)^2.
\label{eq:b_synch}
\end{equation}
where 
\begin{equation}
B_{\text{avg}} = \frac{1}{r_h} \int_0^{r_h} B(r) \, dr.
\label{eq:avg_B}
\end{equation}

When $\widehat{G}$ is close
to 1 and spatial diffusion can be neglected. In the case of LMC, we need to numerically calculate the integration of the Green Function, which has been presented in Ref.\citep{McDaniel2017} by using the RX-DMFIT code \footnote {\it https://github.com/alex-mcdaniel/RX-DMFIT}. We will follow this process and use RX-DMFIT to deal with diffusion equation.

\subsection{Synchrotron emission related to DM annihilation}

\begin{table*}[tbp]
\centering
\begin{tabular}{|c|c|c|c|c|c|c|}
\hline
  $r_h$ (kpc)  & D (kpc)    &   $D_0$ (cm$^2$s$^{-1})$     &      $B_0$ ($\mu$G)    &         $\rho_s$ (GeV/cm$^3$)   &    $r_s$ (kpc)  \\
\hline
 $3.5$ & 50.1 &$3.0\times 10^{27}$& $5.0$ & $0.31$ &$9.04$  \\
\hline
\end{tabular}
\caption{\label{tab:LMC} Parameters in our benchmark model.}
\end{table*}

The synchrotron emission can be approximated as \citep{McDaniel2017}:
\begin{equation}\label{eq:ssyn}
S_{syn} \approx \frac{1}{D^2} \int_{0}^{r_{max}} dr r^2 j_{syn}(\nu, r ),
\end{equation}
where $D= 50.1$ kpc is the distance to LMC.  The value of $r_{max}$ is set to be 3.5 kpc, which enable the emission region is comparable to $8^{\circ}\times8^{\circ}$ images centred on the LMC. As in Ref.\cite{Storm2017}, the synchrotron emissivity is
\begin{equation}
j_{syn} (\nu, r)= 2\int_{m_e}^{M_{\chi}} dE \frac {d{n_{e}}}{dE}
( E, r) P_{syn} ( \nu, E, r )
\end{equation}
The average synchrotron power supply in all directions for a certain frequency $\nu$ is
\begin{equation}
P_{syn} \left(\nu, E , r\right) = \int_0^{\pi} d\theta \frac{\sin \theta}{2} 2\pi \sqrt{3}r_0 m_e c \nu_0 \sin\theta F\left(\frac{x}{\sin\theta}\right),
\end{equation}
where $r_0 = e^2/(m_ec^2)$ represents the classical electron radius, $\theta$ represents the pitch angle, and $\nu_0 = eB/(2\pi m_e c)$ denotes the non-relativistic gyro-frequency. The $x$ and $F$ quantities are defined follows:
\begin{equation}
x \equiv \frac{2\nu \left(1+z\right)m_e^2}{3\nu_0 E^2},
\end{equation}
\begin{equation}
F(s) \equiv s\int_s^{\infty} d \zeta K_{5/3}\left( \zeta \right)\approx 1.25 s^{1/3}e^{-s}\left[648 + s^2\right]^{1/12},
\end{equation}
where $K_{5/3}$ is the modified Bessel function of order 5/3.

\subsection{Constraints on $m_{\chi}$ and ${\langle \sigma v \rangle}$ parameter space}

As illustrated in the above calculation, the synchrotron flux at a given frequency $\nu$ is then determined totally by $m_{\chi}$ and ${\langle \sigma v \rangle}$, which means that the deduced upper limits on synchrotron emission related to DM annihilation at each frequency in last section could draw a corresponding upper limit curve in the $m_{\chi}$ and ${\langle \sigma v \rangle}$ parameter space.

\begin{figure*}
\begin{minipage}[h]{0.49\linewidth}
\center{\includegraphics[width=1 \linewidth]{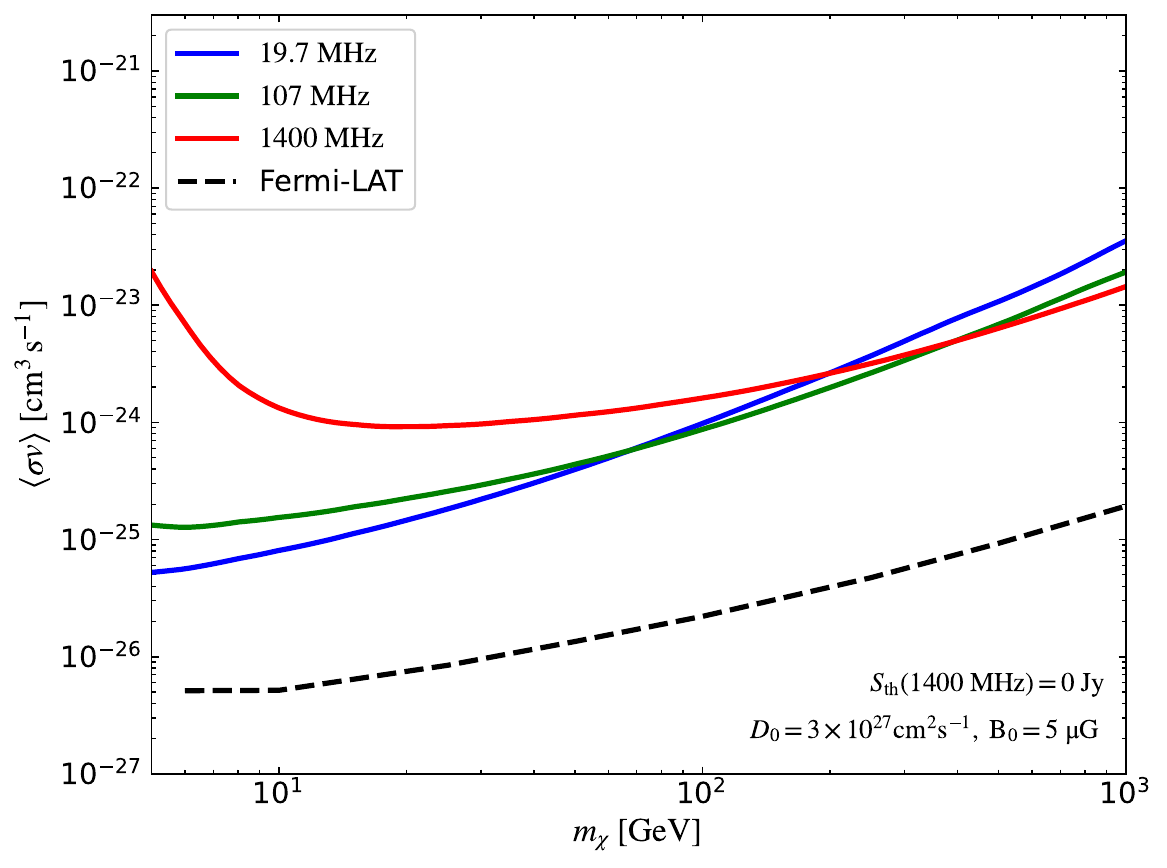}} \\
\end{minipage}
\hfill
\begin{minipage}[h]{0.49\linewidth}
\center{\includegraphics[width=1 \linewidth]{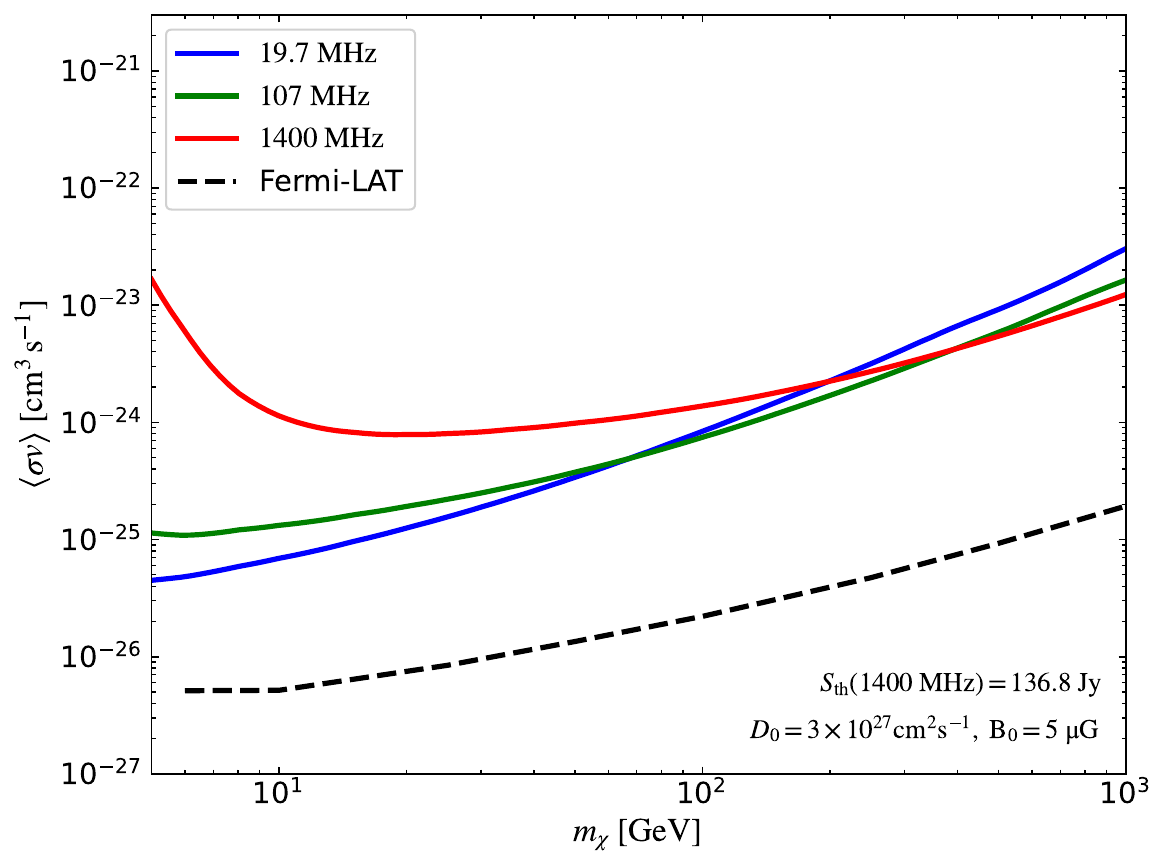} \\ }
\end{minipage}
\caption{Constraints on $m_{\chi}$ and ${\langle \sigma v \rangle}$ parameter space of different frequencies without (left panel) and with (right panel)  thermal radiation.}
\label{LMCch}
\end{figure*}

Utilizing deduced upper limit synchrotron emission related DM annihilation  $S_{\mathrm{DM}}=114.8(\nu/\rm 1.4\,GHz)^{-0.75}$ (without considering thermal radiation ) and $S_{\mathrm{DM}}=98.4 (\nu/\rm 1.4\,GHz)^{-0.75}$ (considering thermal radiation ), we display constraints on the $m_{\chi}$ and $\langle \sigma v \rangle$ parameter space for three frequencies (19.7\,MHz, 107\,MHz, and 1.4\,GHz) in Fig.~\ref{LMCch}. In Table \ref{tab:LMC}, we list parameters in our fiducial model and refer to solid lines in Fig.\ref{LMCch}. When including thermal radiation, the derived constraints are approximately 1.2 times stronger than those obtained without considering thermal radiation. Low-frequency observations impose stronger constraints on low-mass dark matter compared to high-frequency observations.  However, such kinds of constraints are hardly comparable to those obtained from Fermi-LAT gamma-ray\citep{Ackermann2015}. In our benchmark model, all the upper limits are well above the thermal relic cross section ($2.2\times10^{-26}\,\mathrm{cm}^3\,\mathrm{s}^{-1}$ at $m_\chi = 100\,\mathrm{GeV}$), the classical annihilation cross section required for WIMP to make up the dark matter through a thermal production. Numbers of $e^+e^-$ have been significantly reduced in the inner region of the LMC due to diffusion. Deep observation conducted by a low-frequency radio telescope with a large field view will be potential to improve its competitiveness.

\begin{figure*}
\begin{minipage}[h]{0.49\linewidth}
\center{\includegraphics[width=1 \linewidth]{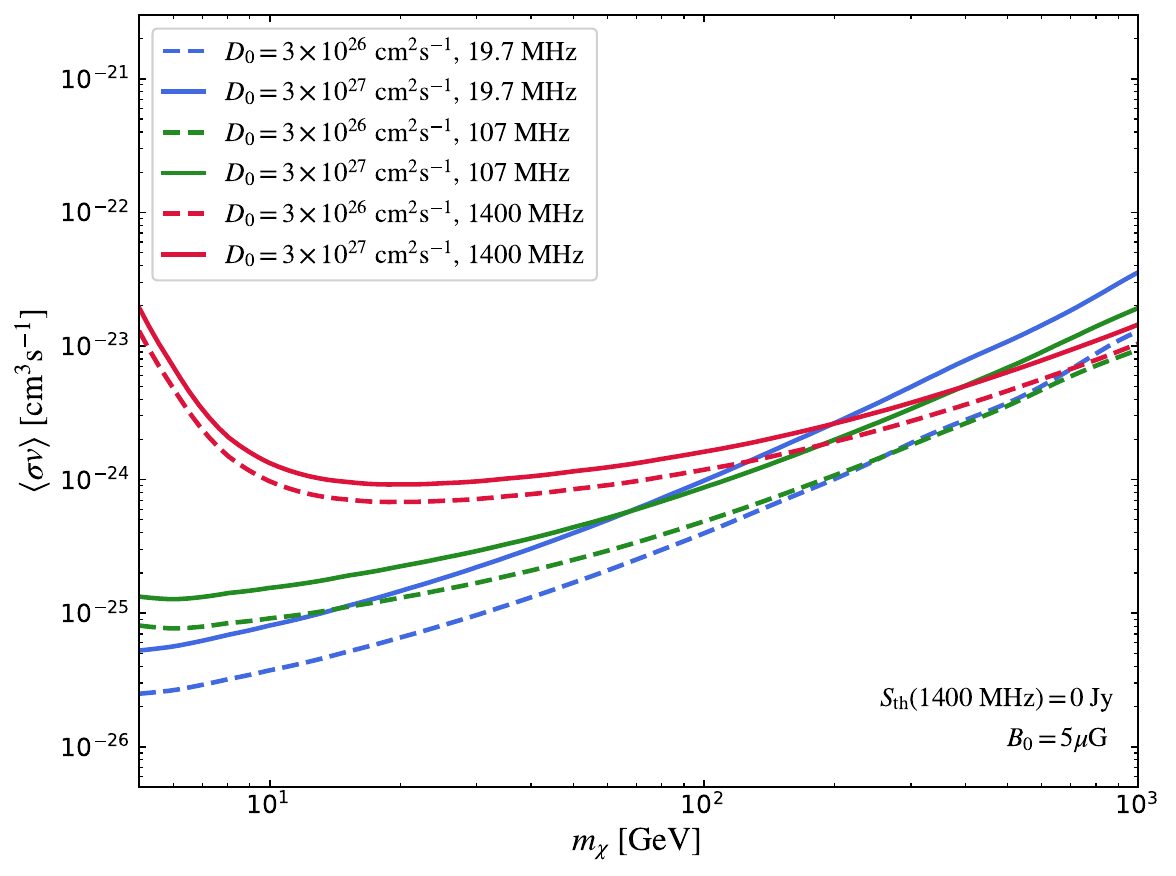} \\ }
\end{minipage}
\hfill
\begin{minipage}[h]{0.49\linewidth}
\center{\includegraphics[width=1 \linewidth]{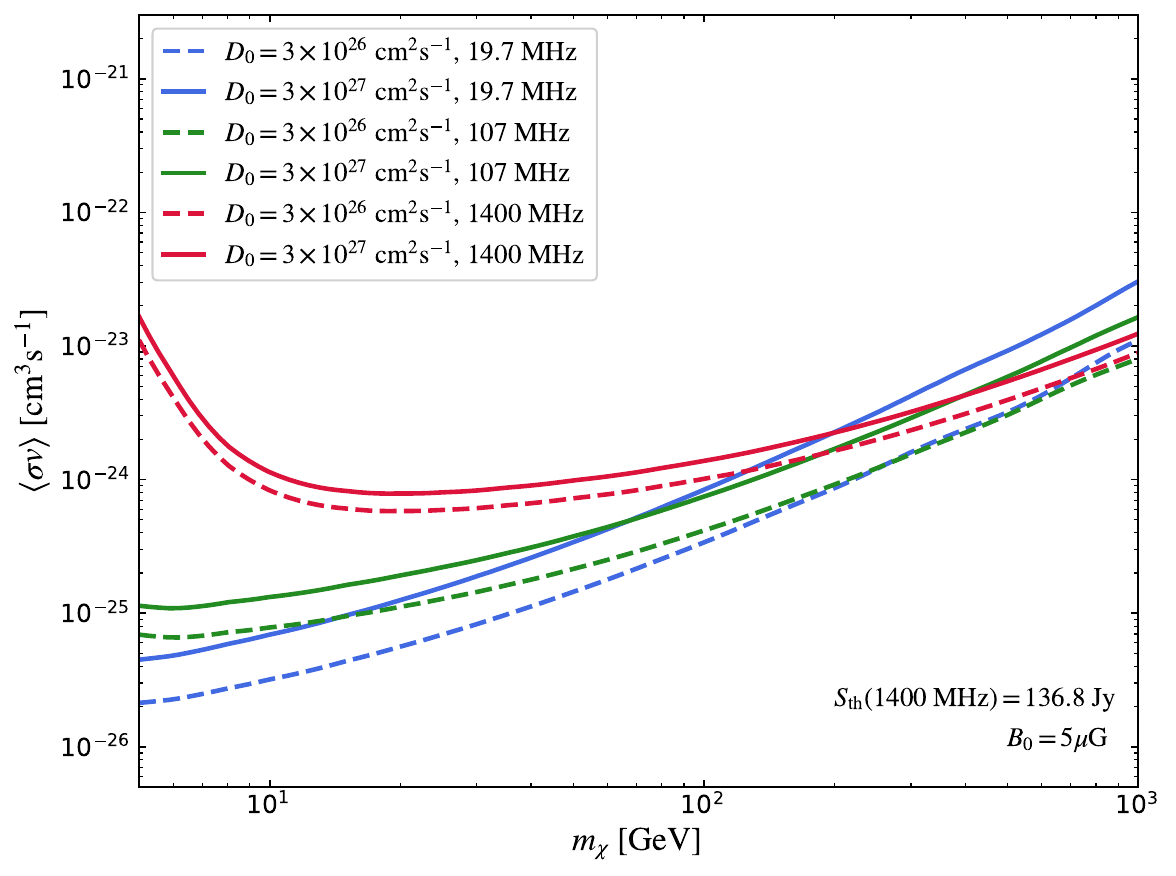}\\} 
\end{minipage}
\caption{Constraints on $m_{\chi}$--$\langle \sigma v \rangle$ are shown for $D_0 = 3 \times 10^{27}~\mathrm{cm}^{2}~\mathrm{s}^{-1}$ (solid) and $3 \times 10^{26}~\mathrm{cm}^{2}~\mathrm{s}^{-1}$ (dashed), where left panel exclude and right panel include thermal radiation. }
\label{LMCD0}
\end{figure*}

The diffusion coefficient of cosmic ray propagation remains uncertain.  Considering that the gravitational potential well of the LMC is weaker than the Milky Way's, we adopt a benchmark value of $D_0 = 3 \times
10^{27}$~cm$^{2}$~s$^{-1}$.   Figs.~\ref{LMCD0} compare $D_0 = 3 \times 10^{26}\,\mathrm{cm}^2\,\mathrm{s}^{-1}$ constraints (dashed) to benchmark, showing weaker diffusion both without (left) and with (right) thermal radiation. When $D_{0}$ is small,  relativistically charged particles with minimal energy loss  escape from the diffusion zone. Our findings indicate stronger constraints under weaker diffusion assumptions.

\begin{figure*}
\begin{minipage}[h]{0.49\linewidth}
\center{\includegraphics[width=1 \linewidth]{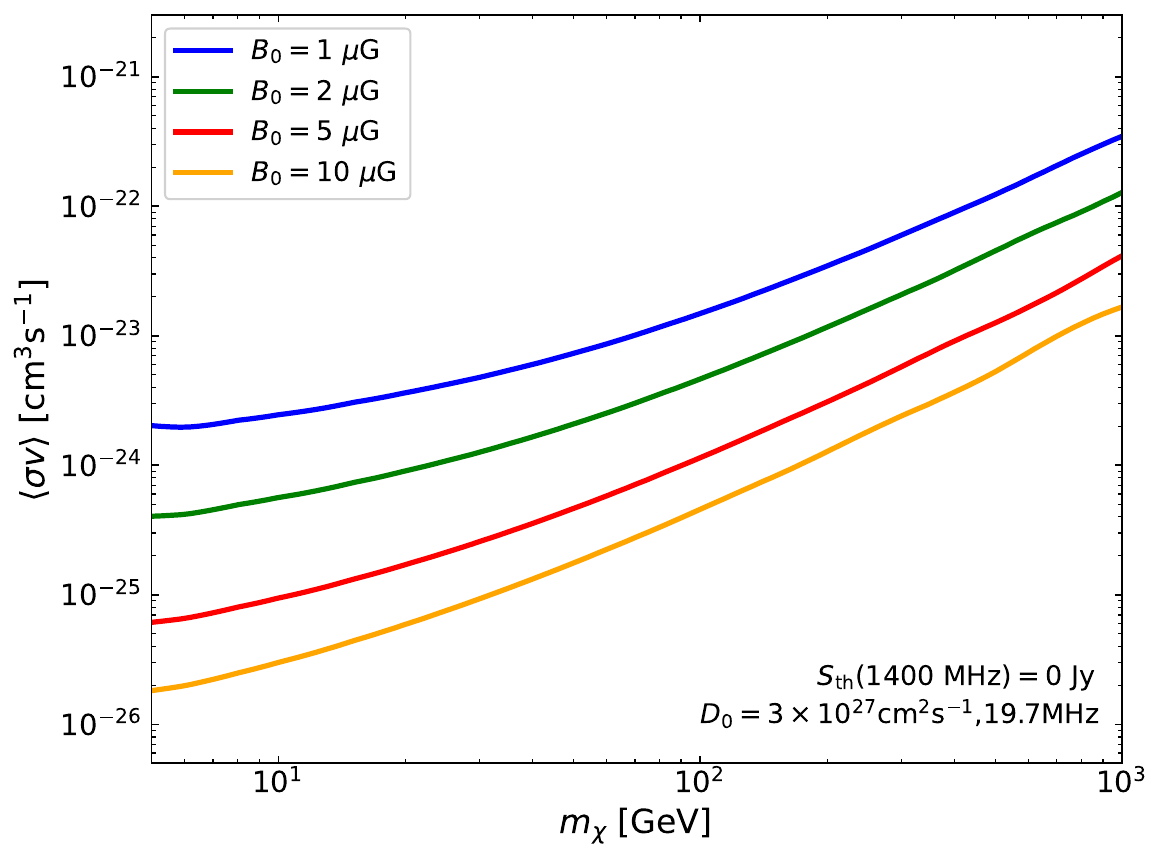} \\ }
\end{minipage}
\hfill
\begin{minipage}[h]{0.49\linewidth}
\center{\includegraphics[width=1 \linewidth]{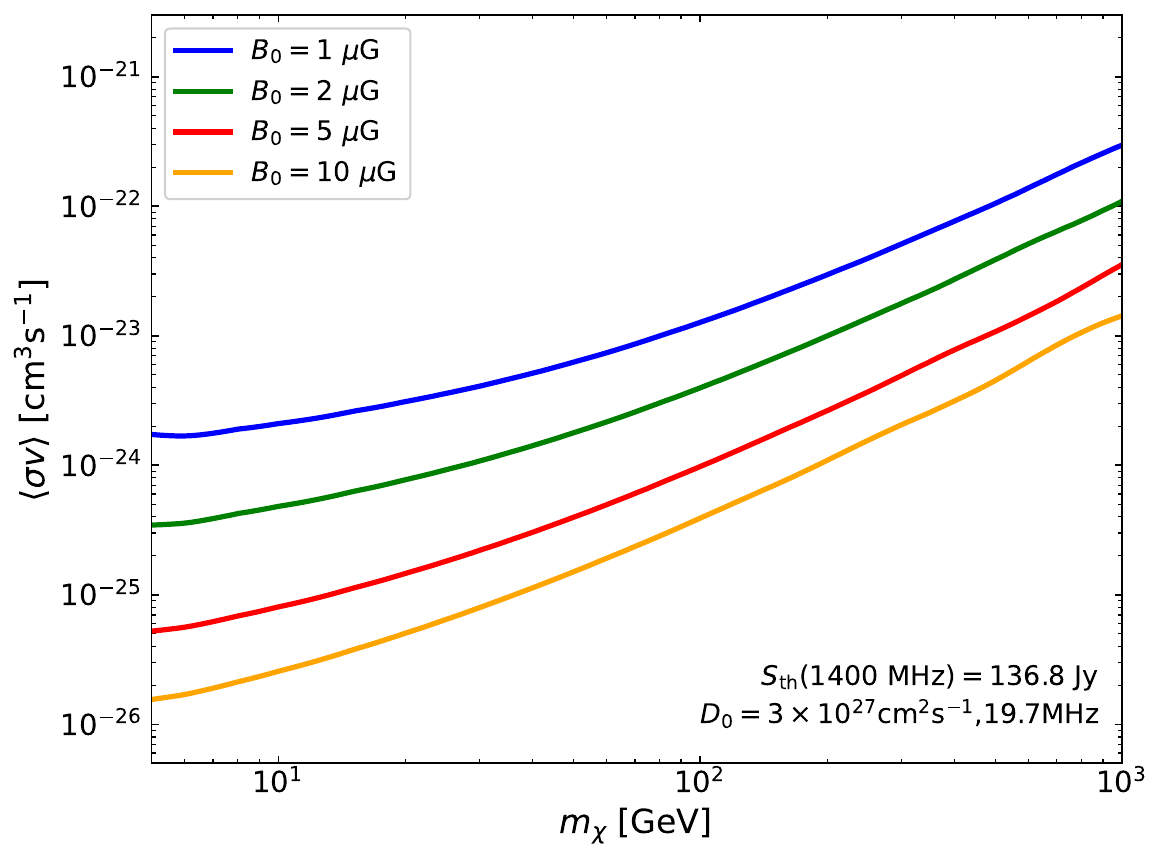} \\ }
\end{minipage}
\caption{Constraints on the $m_{\chi}$--$\langle \sigma v \rangle$ parameter space for magnetic field strengths varying from $1~\mu\mathrm{G}$ to $10~\mu\mathrm{G}$ with left panel excluding and right panel including thermal radiation contributions. }
\label{LMCB0}
\end{figure*}

The magnetic field is the most crucial factor in calculating the synchrotron emission \cite{Chi1993}. However, the detailed magnetic field distribution in LMC is far from clear. As for the estimated total magnetic field, some studies converge roughly to 6 $\mu G$(\cite{Klein1989,Haynes1991}). The maximum value of the total magnetic field is approximately equal to  $ 18.4~ \mu G$. The averaged strength of the magnetic field in the inner region could be as high as $5 ~\mu G$. We conservatively adopt the maximum central LMC magnetic field strength $\sim 5\,\mu\mathrm{G}$, shown by red solid lines in Fig.~\ref{LMCB0}'s left (no thermal) and right (with thermal) panel. We also illustrate the constraints while changing magnetic field strength from $1 ~\mu G$ to $10 ~\mu G$. It is clear that the deduced upper limits on DM annihilation cross section are sensitive to the assumed magnetic fields.

\section{Summary}
In this paper, we present revised constraints on dark matter particle properties by fitting the updated multiple low-frequency radio observations in LMC. Meanwhile, we take into account two different sources of energetic $e^+/e^-$: dark matter annihilation and cosmic rays. For the CR component, the best-fit values are $\alpha_{\mathrm{CR}} = 0.40$ and $S_{\mathrm{CR}}(1.4~\mathrm{GHz}) = 354.8~\mathrm{Jy}$ in the case without a thermal component, and $\alpha_{\mathrm{CR}} = 0.52$ and $S_{\mathrm{CR}}(1.4~\mathrm{GHz}) = 239.9~\mathrm{Jy}$ in the case with a thermal component. These normalization factors are higher than those predicted by the radio/far-infrared relation based on observations of normal galaxies. Moreover, the fitted values of $\alpha_{\mathrm{CR}}$ are lower than the canonical value of $\alpha_{\mathrm{CR}} \approx 0.80$ commonly adopted in the literature, indicating a flatter spectrum than typically assumed for normal galaxies. Depending on whether thermal emission is included or not, our best-fit values for $S_{\mathrm{DM}}(\nu_{\star})$ range from 98.4 Jy to 114.8 Jy. Low-frequency observations yield significantly tighter constraints on low-mass dark matter compared to high-frequency observations. Our analysis reveals two trends: (i) weaker diffusion conditions yield stronger limits, and (ii) higher magnetic field intensities produce more stringent constraints, as shown in Fig.~\ref{LMCD0} and Fig.~\ref{LMCB0}. Future extremely low-frequency radio surveys, such as the future low-frequency Square Kilometre Array (SKA) and the proposed radio interferometer array in space, should be considered as a promising and powerful way to constrain dark matter.


\section{Acknowledgements}

We thank Renyi Ma for his help using emcee code. We thank Hai-kun Li for his support during the preparation for this manuscript. This work is supported by the National Natural Science Foundation of China under Nos. 11890692, 12133008, 12221003. We acknowledge the science research grant from the China Manned Space Project with No. CMS-CSST-2021-A04.


\begin{thebibliography}{99}


\bibitem{Feng:2010gw}
J.~L.~Feng,
Ann. Rev. Astron. Astrophys. \textbf{48}, 495-545 (2010)
doi:10.1146/annurev-astro-082708-101659
[arXiv:1003.0904 [astro-ph.CO]].

\bibitem[Porter et al.(2011)]{reviewII} Porter, T.A., Johnson, R.P., and Graham, P.W.: 2011, Annual Review of Astronomy and Astrophysics, 49, 155. doi:10.1146/annurev-astro-081710-102528.

\bibitem[Bertone et al.(2005)]{WIMPII} Bertone G., Hooper D., Silk J.  \ 2005, Phys. Rept.
405, 279


\bibitem[Jungman et al.(1996)]{WIMPI} Jungman G., Kamionkowski M., Griest K.\ 1996, Phys. Rept.
267, 195


\bibitem{ColafrancescoATCTI}
M.~Regis, L.~Richter, S.~Colafrancesco, M.~Massardi, W.~J.~G.~de Blok, S.~Profumo and N.~Orford,
Mon. Not. Roy. Astron. Soc. \textbf{448}, no.4, 3731-3746 (2015)
doi:10.1093/mnras/stu2747
[arXiv:1407.5479 [astro-ph.GA]].

\bibitem{ColafrancescoATCTII}
M.~Regis, L.~Richter, S.~Colafrancesco, S.~Profumo, W.~J.~G.~de Blok and M.~Massardi,
Mon. Not. Roy. Astron. Soc. \textbf{448}, no.4, 3747-3765 (2015)
doi:10.1093/mnras/stv127
[arXiv:1407.5482 [astro-ph.GA]].

\bibitem[Regis et al.(2014)]{ColafrancescoATCTIII} Regis, M., Colafrancesco, S., Profumo, S., de Blok, W. J. G.,
Massardi, M., Laura, R., \ 2014, \JCAP, 10, 016

\bibitem[Regis et al.(2017)]{ColafrancescoATCTReti} Regis, M., Laura, R., Colafrancesco \ 2017, \JCAP, 07, 025

\bibitem[Kristine et al.(2013)]{GBTI}Spekkens, K., Laura, R., Mason, B. S., Aguirre, J. E., Nhan, B \ 2013, \apj, 773, 61 

\bibitem[Natarajan et al.(2013)]{GBTII}Natarajan, A., Peterson, J. B., Voytek, T. C., Spekkens, K., Mason, B., Aguirre, J.,  Willman, B.\ 2013, \prd, 88, 3535


\bibitem[Kar et al.(2019)]{MWA2019} Kar A., Mitra S., Mukhopadhyaya B., Choudhury,T. R.  \ 2019, Phys. Rev. D 100, 043002


\bibitem[Basu et al.(2021)]{Basu2021} Basu, A., Roy, N., Choudhuri, S., Datta, K.K., and Sarkar, D.: 2021,  Monthly Notices of the Royal Astronomical Society, 502, 1605. doi:10.1093/mnras/stab120.


\bibitem{Guo2022} 
W.~Q.~Guo, Y.~Li, X.~Huang, Y.~Z.~Ma, G.~Beck, Y.~Chandola and F.~Huang,
Phys. Rev. D \textbf{107}, no.10, 103011 (2023)
doi:10.1103/PhysRevD.107.103011
[arXiv:2209.15590 [astro-ph.HE]].

\bibitem[Sofue(1999)]{Sofue1999} Sofue, Y.: 1999, Publications of the Astronomical Society of Japan,51 , 445. doi:10.1093/pasj/51.4.445.

\bibitem{Gammaldi:2021zdm}
V.~Gammaldi \textit{et al.} [Fermi-LAT],
PoS \textbf{ICRC2021}, 509 (2021)
doi:10.22323/1.395.0509
[arXiv:2109.11291 [astro-ph.CO]].



\bibitem{Regis2021}
M.~Regis, J.~Reynoso-Cordova, M.~D.~Filipovi\'c, M.~Br\"uggen, E.~Carretti, J.~Collier, A.~M.~Hopkins, E.~Lenc, U.~Maio and J.~R.~Marvil, \textit{et al.}
JCAP \textbf{11}, no.11, 046 (2021)
doi:10.1088/1475-7516/2021/11/046
[arXiv:2106.08025 [astro-ph.HE]].
\bibitem{Salucci2000ps}
P.~Salucci and A.~Burkert,
Astrophys. J. Lett. \textbf{537}, L9-L12 (2000)
doi:10.1086/312747
[arXiv:astro-ph/0004397 [astro-ph]].

\bibitem[Siffert et al.(2011)]{Siffert2011} Siffert, B.B., Limone, A., Borriello, E., Longo, G., and Miele, G.: 2011,  Monthly Notices of the Royal Astronomical Society,410, 2463. doi:10.1111/j.1365-2966.2010.17613.x.
\bibitem[Klein et al.(1989)]{Klein1989} Klein, U., Wielebinski, R., Haynes, R.F., and Malin, D.F.: 1989,  Astronomy and Astrophysics, 211, 280.

\bibitem[Alvarez et al.(1987)]{Alvarez1987} Alvarez, H., Aparici, J., and May, J.: 1987,  Astronomy and Astrophysics, 176, 25.

\bibitem[Mills(1959)]{Mills1959} Mills, B.~Y.\ 1959, Handbuch der Physik, 53, 239 

\bibitem[Shain(1959)]{Shain1959} Shain, C.~A.\ 1959, URSI Symp.~1: Paris Symposium on Radio Astronomy, 9, 328


\bibitem[For et al.(2018)]{For2018} 
For, B.-Q., Staveley-Smith, L., Hurley-Walker, N., Franzen, T., Kapi{\'n}ska, A.D., Filipovi{\'c}, M.D., and, ...: 2018,  Monthly Notices of the Royal Astronomical Society, 480, 2743. doi:10.1093/mnras/sty1960.

\bibitem[Tasitsiomi et al.(2004)]{Tasitsiomi2004} Tasitsiomi, A., Gaskins, J., \& Olinto, A.~V.\ 2004, Astroparticle Physics, 21, 637

\bibitem[Chan \& Lee(2022)]{Chan2022} Chan, M.~H. \& Lee, C.~M.\ 2022, \apj, 933, 130. doi:10.3847/1538-4357/ac71a9


\bibitem[Haynes et al.(1991)]{Haynes1991} Haynes, R.F., Klein, U., Wayte, S.R., Wielebinski, R., Murray, J.D., Bajaja, E., and, ...: 1991,  Astronomy and Astrophysics, 252, 475.




\bibitem[Condon(1992)]{Condon1992} Condon, J.~J.\ 1992, Annual Rev. Astron. Astrophys, 30, 575

\bibitem[Berezhko(2014)] {Berezhko2014} Berezhko, E.~G.\ 2014, Nuclear Physics B Proceedings Supplements, 256, 23

\bibitem[Rieke et al.(2009)]{Rieke2009} Rieke, G.~H., Alonso-Herrero, A., Weiner, B.~J., et al.\ 2009, \apj, 692, 556


\bibitem[Lawton et al.(2010)]{Lawton2010} Lawton, B., Gordon, K.~D., Babler, B., et al.\ 2010, \apj, 716, 453


\bibitem[Foreman-Mackey et al.(2016)]{Foreman2016} Foreman-Mackey, D., Vousden, W., Price-Whelan, A., et al.\ 2016, Zenodo Software Release, 2016


\bibitem[Hassani et al.(2022)]{Hassani2022}Hassani, H., Tabatabaei, F., Hughes, A., et al.\ 2022,  Monthly Notices of the Royal Astronomical Society, 510, 11. doi:10.1093/mnras/stab3202.

\bibitem{Navarro1995iw}
J.~F.~Navarro, C.~S.~Frenk and S.~D.~M.~White,
Astrophys. J. \textbf{462}, 563-575 (1996)
doi:10.1086/177173
[arXiv:astro-ph/9508025 [astro-ph]].

\bibitem{Moore1999gc}
B.~Moore, T.~R.~Quinn, F.~Governato, J.~Stadel and G.~Lake,
Mon. Not. Roy. Astron. Soc. \textbf{310}, 1147-1152 (1999)
doi:10.1046/j.1365-8711.1999.03039.x
[arXiv:astro-ph/9903164 [astro-ph]].


\bibitem{Hayashi2002qv}
E.~Hayashi, J.~F.~Navarro, J.~E.~Taylor, J.~Stadel and T.~R.~Quinn,
Astrophys. J. \textbf{584}, 541-558 (2003)
doi:10.1086/345788
[arXiv:astro-ph/0203004 [astro-ph]].

\bibitem{Bonnivard2015}
V.~Bonnivard, C.~Combet, M.~Daniel, S.~Funk, A.~Geringer-Sameth, J.~A.~Hinton, D.~Maurin, J.~I.~Read, S.~Sarkar and M.~G.~Walker, \textit{et al.}
Mon. Not. Roy. Astron. Soc. \textbf{453}, no.1, 849-867 (2015)
doi:10.1093/mnras/stv1601
[arXiv:1504.02048 [astro-ph.HE]].

\bibitem[Colafrancesco et al.(2006)]{diffuse} Colafrancesco, S., Profumo, S. \& Ullio, P.\ 2006, Astron. Astrophys. 455, 21

\bibitem{Maurin:2001sj}
D.~Maurin, F.~Donato, R.~Taillet and P.~Salati,
Astrophys. J. \textbf{555} (2001), 585-596
doi:10.1086/321496
[arXiv:astro-ph/0101231 [astro-ph]].

\bibitem{Webber:1992dks}
W.~R.~Webber, M.~A.~Lee and M.~Gupta,
Astrophys. J. \textbf{390} (1992), 96
doi:10.1086/171262

\bibitem{Weingartner2001}
J.~C.~Weingartner and B.~T.~Draine,
Astrophys. J. Suppl. \textbf{134}, 263-282 (2001)
doi:10.1086/320852
[arXiv:astro-ph/9907251 [astro-ph]].

\bibitem[McDaniel et al.(2017)]{McDaniel2017} McDaniel, A., Jeltema, T., Profumo, S., \& Storm, E.\ 2017, JCAP, 9, 027

\bibitem[Cholis et al.(2015)]{Cholis2015} Cholis, I., Hooper, D., \& Linden, T.\ 2015, \prd, 91, 083507

\bibitem[Storm et al.(2017)]{Storm2017} Storm, E., Jeltema, T.~E., Splettstoesser, M., \& Profumo, S.\ 2017, \apj, 839, 33

\bibitem[Ackermann et al.(2015)]{Ackermann2015} Ackermann, M., Albert, A., Anderson, B., et al.\ 2015, Physical Review Letters, 115, 231301

\bibitem[Chi \& Wolfendale(1993)]{Chi1993} Chi, X., \& Wolfendale, A.~W.\ 1993, \nat, 362, 610




































\end{thebibliography}
\end{document}